\documentclass[fleqn,usenatbib]{mnras}

\usepackage{newtxtext,newtxmath}

\usepackage[T1]{fontenc}

\DeclareRobustCommand{\VAN}[3]{#2}
\let\VANthebibliography\thebibliography
\def\thebibliography{\DeclareRobustCommand{\VAN}[3]{##3}\VANthebibliography}

\usepackage{graphicx}	
\usepackage{amsmath}	
\usepackage{newtxtext,newtxmath}
\usepackage{indentfirst}
\usepackage{tabularx}
\usepackage{array}
\usepackage{subfigure}
\usepackage{graphics}
\usepackage[justification=centering]{caption}
\usepackage{dcolumn}
\usepackage{bm}
\usepackage{color}
\usepackage{diagbox}
\usepackage{amsmath}
\usepackage{cancel}
\usepackage[draft,defaultcolor=red]{changes}

\title[Double Hall pattern in dusty plasmas]{Discovery of double Hall pattern associated with collisionless magnetic reconnection in dusty plasmas}

\author[S.-D. Yang, L. Wang, and C. Dong]{
Shu-Di Yang,$^{1}$\thanks{E-mail: sdyang@pku.edu.cn}
Liang Wang,$^{2}$\thanks{E-mail: lw6@princeton.edu}
and Chuanfei Dong$^{2,3}$\thanks{E-mail: dcfy@bu.edu}
\\
$^{1}$School of Physics, Peking University, Beijing 100871, China\\
$^{2}$Princeton Plasma Physics Laboratory and Department of Astrophysical Sciences, Princeton University, Princeton, New Jersey 08540, USA\\
$^{3}$Department of Astronomy, Boston University, Boston, Massachusetts 02215, USA \\
}

\date{Accepted 2023 April 30. Received 2023 April 30; in original form 2022 June 17}

\pubyear{2023}

\begin{document}
\label{firstpage}
\pagerange{\pageref{firstpage}--\pageref{lastpage}}
\maketitle

\begin{abstract}
Magnetic reconnection is prevalent in magnetized plasmas in space and laboratories. Despite significant investigations on reconnection in electron-ion plasmas, studies of reconnection in magnetized plasmas with negatively charged dust grains are quite sparse. Here we report the first fully kinetic simulations of collisionless reconnection in a three-species (i.e., electron, proton, and negatively charged dust grain) dusty plasma, through which the discovery of double Hall pattern is made. The double Hall pattern consists of a traditional Hall quadruple current in between the ion and electron diffusion region, and a reversed Hall current in between the boundary of the ion and dust diffusion region. The analysis of the reconnection rate is also given. This study may be applicable to explain observations of planetary magnetospheres and the astrophysical objects, and may be realized in the laboratory studies of dusty plasmas.
\end{abstract}




\section{Introduction}
\label{intro}
Dusty plasma, owing to its significant role in astrophysical environments such as in the low solar atmosphere, cometary tails \citep{jovanovic2005}, protoplanetary nebulae, discs around young stellar objects \citep{birk1998} and the interstellar medium (ISM), is gradually attracting attention and offering a new paradigm for studies on basic plasma physics. Dusty plasma is usually composed of three species: electrons, ions, and dust grains. Negatively charged dust grains by absorbing electrons are typically micron- and submicron-sized. The charges and masses of dust grains can vary in a large range for different scenarios \citep{yaroshenko2018}, where the dust grains serve as immobile background when it is very heavy or moving particles otherwise. The dust component, therefore, influences the plasma by introducing new modes to waves and instabilities \citep{Merlino1998} as well as modifying the traditional mode properties \citep{Morfill2009, ghosh2002,debnath2020}. Among the pioneer works, the properties of a vast range of dusty plasmas have been addressed, including the dynamics and equilibrium for different processes \citep{mendis1994}, the basic MHD equations as a basis for other research \citep{shukla1996,birk1996}, equilibrium Harris sheet \citep{lazerson2011}, and the laboratory studies of waves and instabilities \citep{Merlino1998}. 

The characteristic parameters of dusty plasmas vary tremendously, ranging from space to interplanetary, astrophysical, and laboratory environments. In general, the most important components in a dusty plasma are electrons, ions, and negatively charged dust grains. The dust number density could be as low as ${10}^{-6}$ cm$^{-3}$ in molecular cloud while reach as high as ${10}^{6}$ cm$^{-3}$ in Magnetized Dusty Plasma Experiment (MDPX). 

Magnetic reconnection is a ubiquitous process in space, astrophysical, and laboratory plasmas \citep{Zweibel2009,Wang2018,Dong2018,Dong2019,Raymond2018,Dong2022,Li2023}. Whereas reconnection in electron-ion plasma and pair plasmas have been immensely explored \citep{yamada2010}, there have not been many investigations on reconnection in dusty plasmas despite its prevalence in space and astrophysical environments. The majority of previous studies on reconnection in dusty plasmas focused on analytic theories in either resistive \citep{jovanovic2005, shukla2002, birk1996} or collisionless \citep{jovanovic2002} environment. It was found that the presence of charged dust grains modifies the properties of plasma waves and instabilities through quasi-neutrality, and the linear instabilities can be categorized into several more regimes due to the scales and parameters introduced by the dust species. 

Due to the limitation of analytic theory, previous studies on dusty reconnection are mainly focused on the growth rate of linear instabilities, without looking into the detailed structure associated with magnetic reconnection. However, the introduction of the dust species definitely leads to new complicated structures. Hall effect, intensively studied in electron-ion plasmas \citep{sonnerup1979,terasawa,shay1998,ren2005}, 
is one of the mechanisms to accelerate the reconnection process and acts on current sheet dynamics. It leads to a characteristic quadruple out-of-plane magnetic field called the Hall field, which is now found to be exotic in dusty plasmas. In some dusty plasmas, the depletion of electrons and the increase of negatively charged dust grains reverse the Hall effect and render the Hall conductivity negative. The reversed Hall effect was found both analytically \citep{yaroshenko2018} and numerically \citep{Kriegel2011}, and is supported by Cassini observations in Enceladus plasma \citep{Simon2011}. It is, therefore, natural to conjecture that when electron and dust grains both play important roles and scale separations are clear among electron, ion, and dust diffusion regions, a complicated double Hall pattern may emerge. To test the conjecture and investigate the properties of collisionless magnetic reconnection in dusty plasmas, we carry out the first fully kinetic simulation for a three-species (i.e., electron, proton, and negatively charged dust grain) dusty plasma and report our findings in this study.

This paper is organized as follows: in Sec. \ref{dust}, we introduce the numerical model and simulation setup; Sec. \ref{hall} presents the double Hall pattern; in Sec. \ref{Rrate}, the associated reconnection rate is discussed with respect to its time evolution, and Sec. \ref{sum} serves as a conclusion and outlook.

\section{Numerical Model and Simulation Setup}
\label{dust}
For simplicity, we consider a three-species, electron-ion-dust plasma in the present work. Each species is characterized by its own constant particle mass ${{m}_{\alpha }}$ and charge ${{q}_{\alpha }}$, $\alpha =e,i,d$. A single dust grain is charged to $Z_d$ as a consequence of the absorbed free electrons. Hence, ${{q}_{i}}=-{{q}_{e}}=e$, ${{q}_{d}}=-{{Z}_{d}}e$. 

The presence of dust grains introduces new spatial and temporal scales. Besides the characteristic scales shaped by electrons and ions, dusty plasmas introduce the inertial length of the dust grain ${{d}_{d}}=c/{{\omega }_{pd}}$, where ${{\omega }_{pd}}={{(4\pi {{n}_{d}}{{e}^{2}}{{Z}^{2}}/{{m}_{d}})}^{1/2}}$, and the dust gyroradius $\sqrt{2{{m}_{d}}{{T}_{d}}}c/(ZeB)$. With other conditions fixed, the heavier the mass of a dust grain, the larger the dust inertial length. Generally, the mass difference between ions and dust grains is much larger than the charge difference. Increasing the density of the dust, however, shrinks the dust inertial length. It is natural to expect that in dusty plasma reconnection, there should be four regions: the outermost MHD region, the dust diffusion region, the ion diffusion region, and the electron diffusion region. These will conceivably increase the complexity of the reconnection process. 

\begin{figure*}
\centering
\subfigure[]{\label{1a}
	\includegraphics[width=0.48\linewidth]{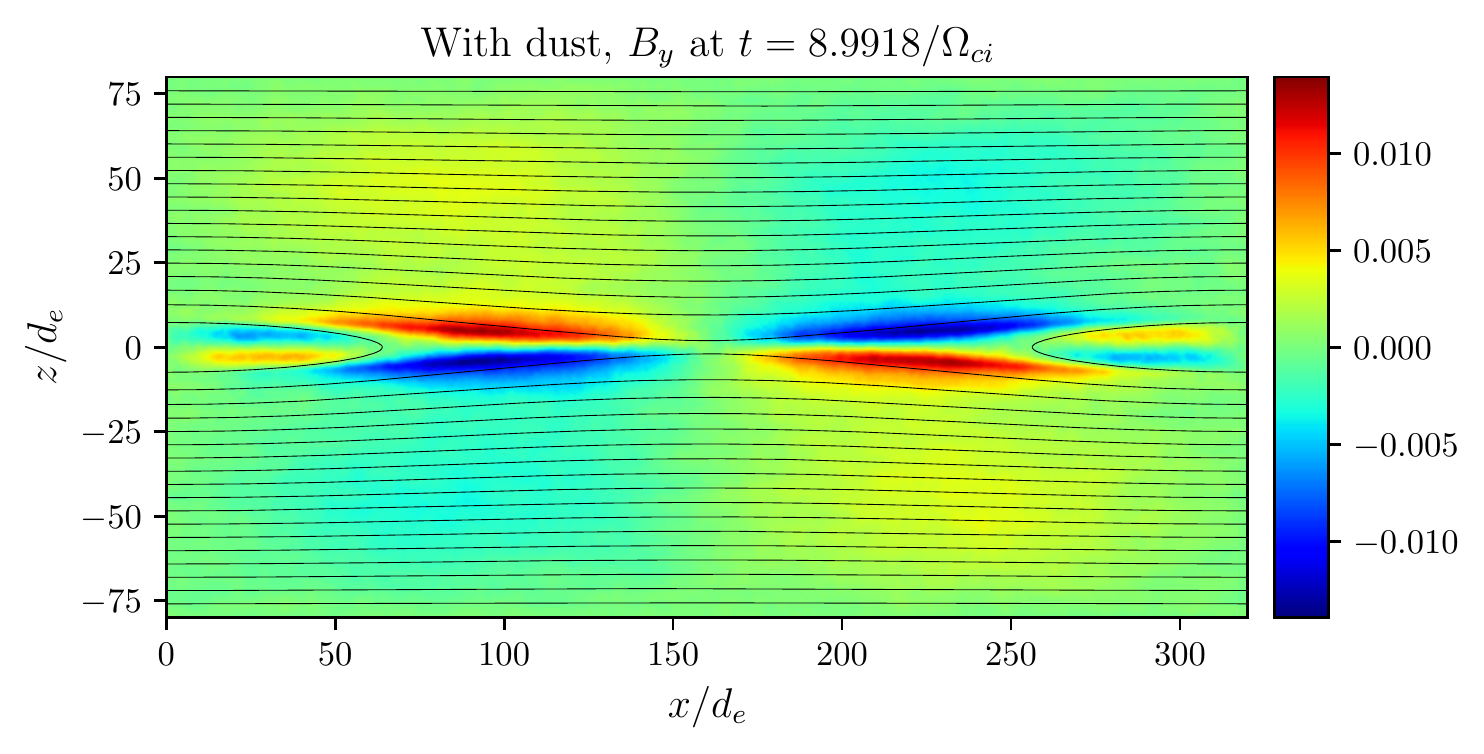}}
\subfigure[]{\label{2a}
	\includegraphics[width=0.48\linewidth]{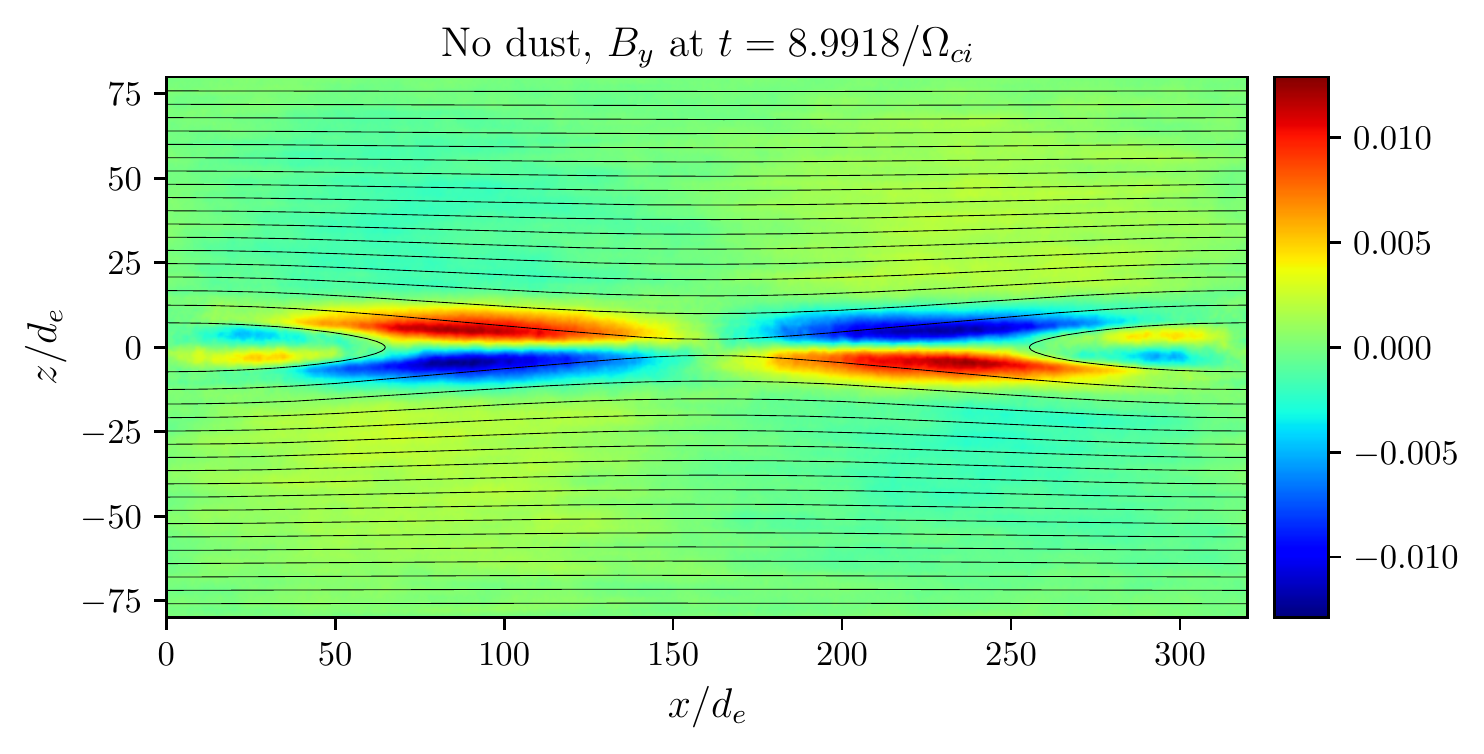}}
\subfigure[]{\label{3a}
	\includegraphics[width=0.48\linewidth]{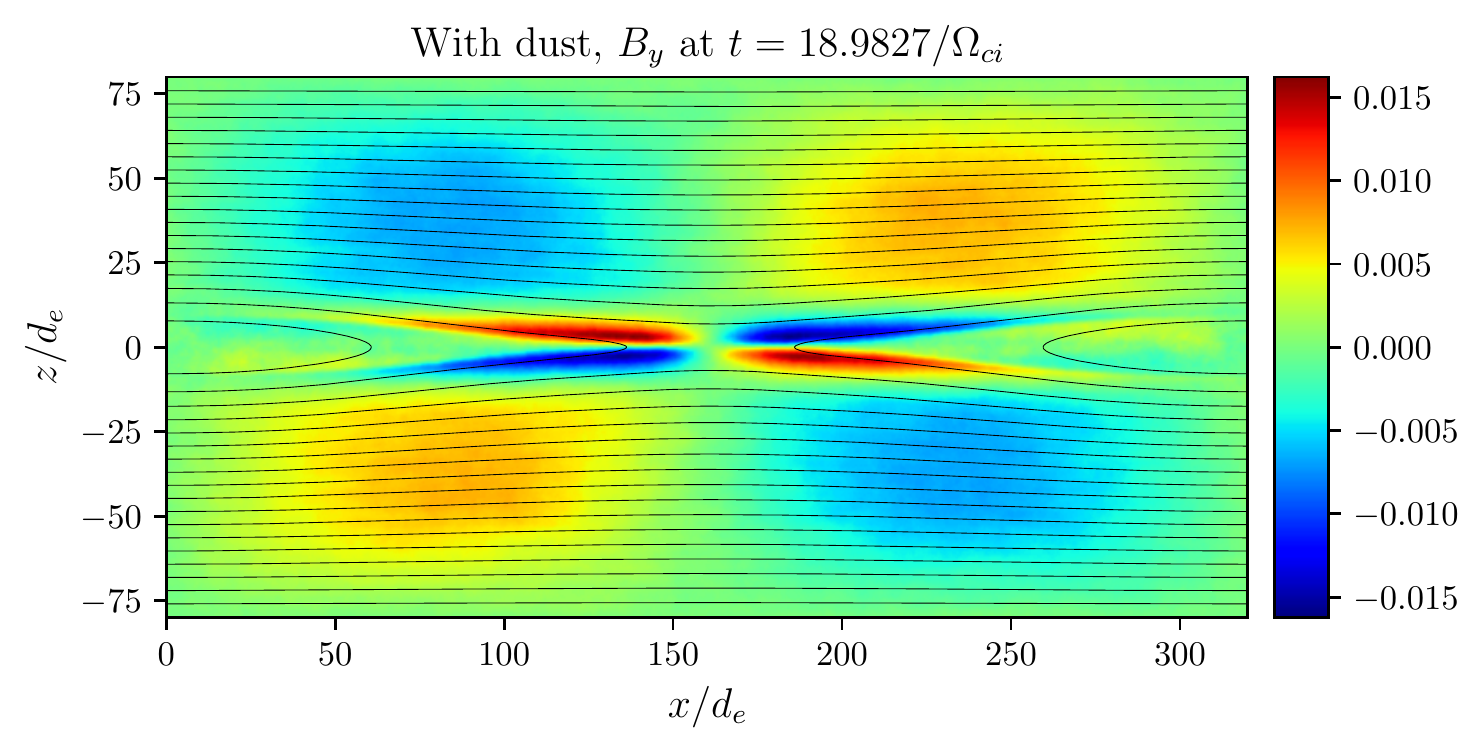}}
\subfigure[]{\label{1b}
	\includegraphics[width=0.48\linewidth]{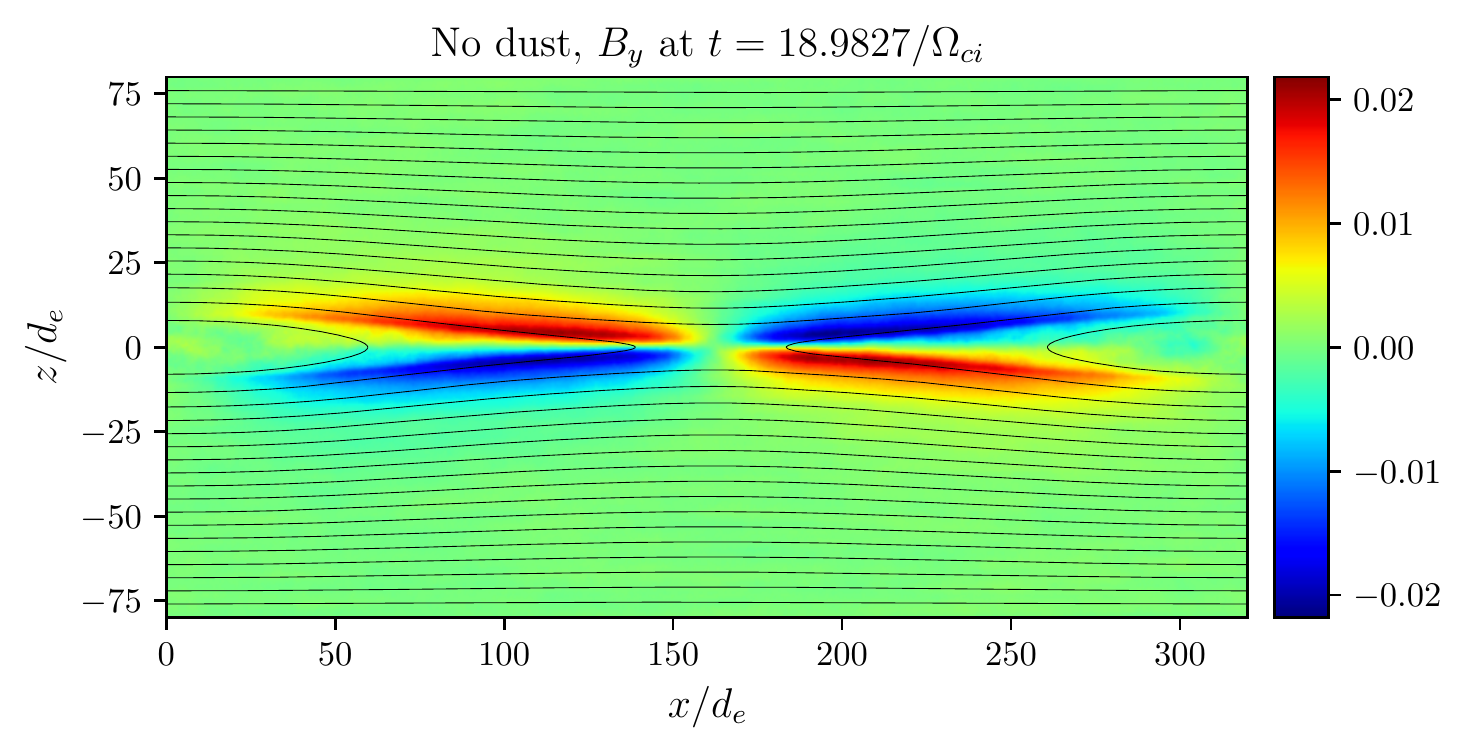}}
\subfigure[]{\label{2b}
	\includegraphics[width=0.48\linewidth]{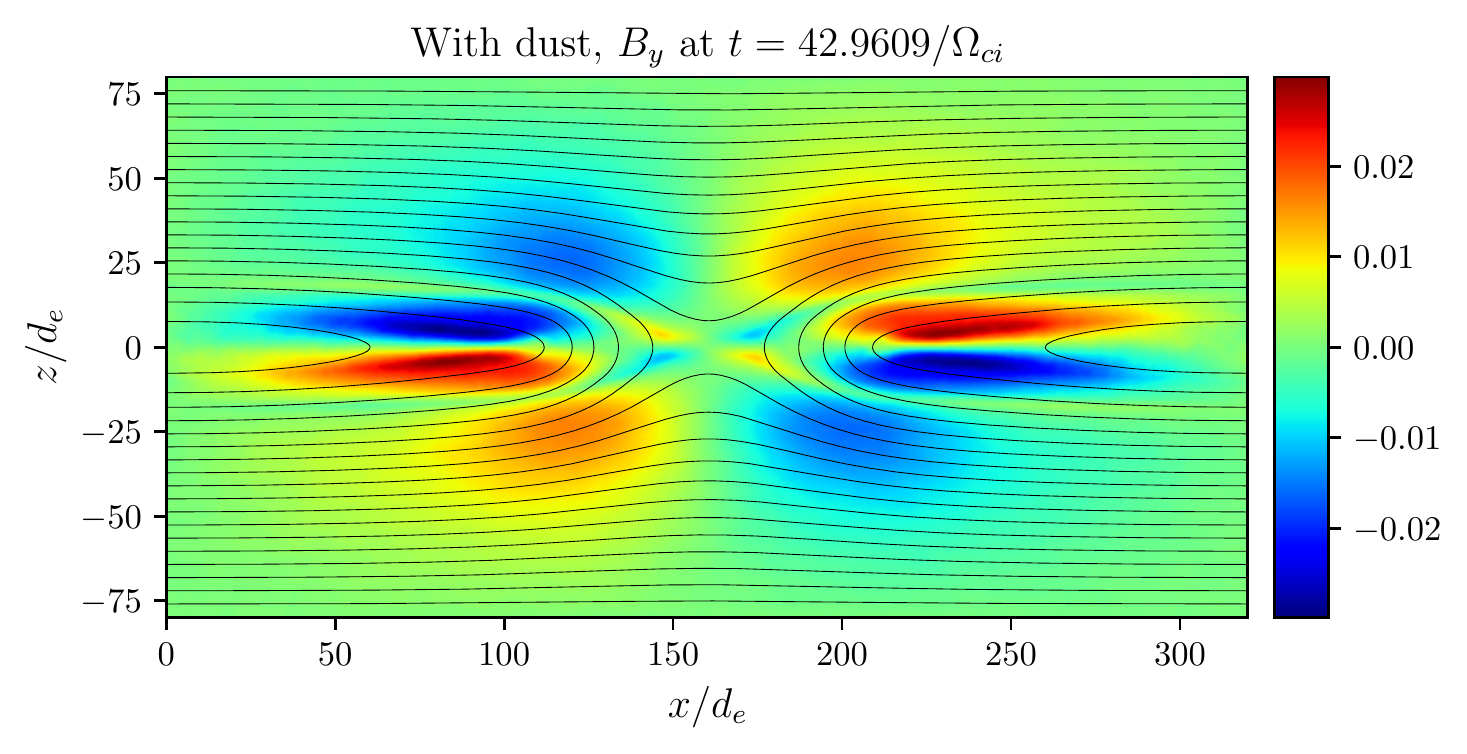}}
\subfigure[]{\label{3b}
	\includegraphics[width=0.48\linewidth]{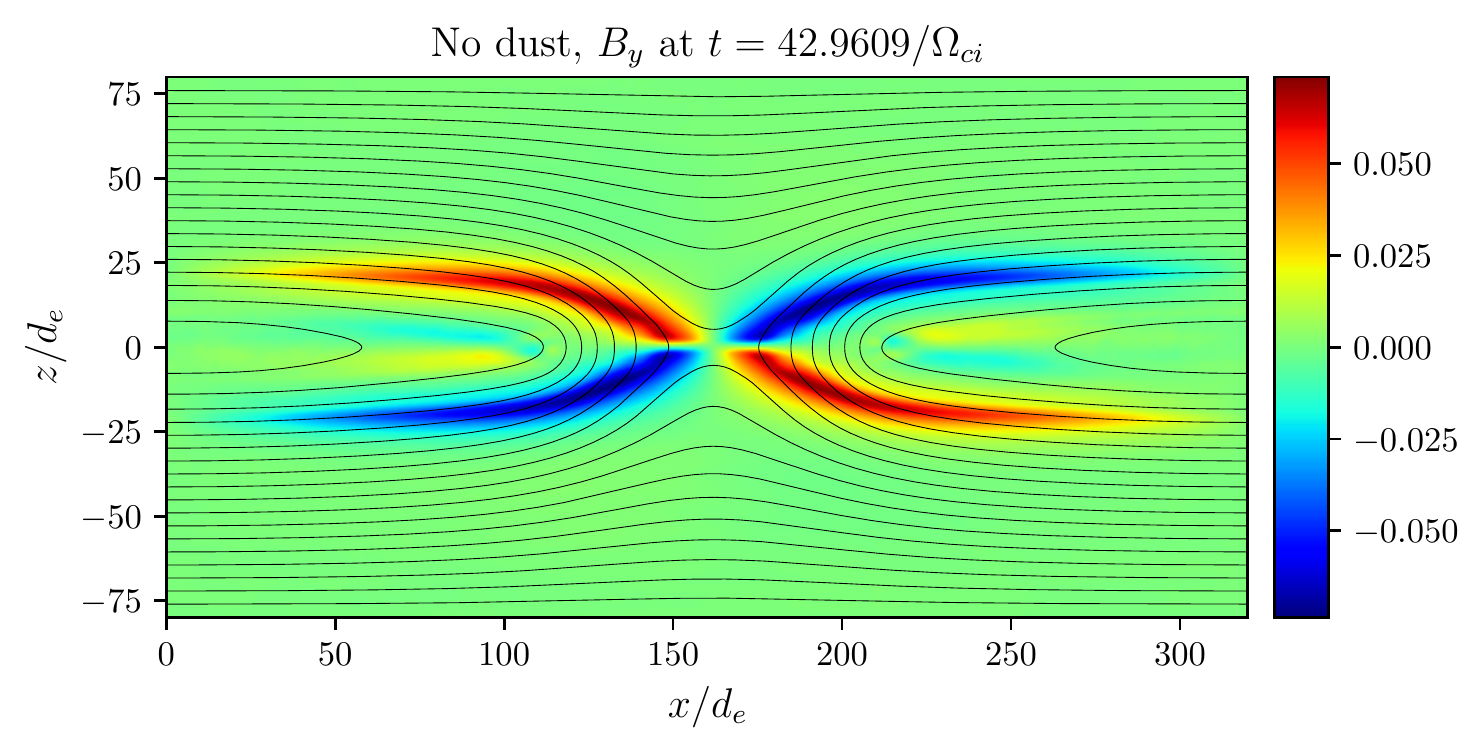}}
    \caption{The simulated out-of-plane magnetic field, $B_y$, associated with collisionless magnetic reconnection with dust grains (left) and without dust grains (right) at three different time steps.}
    \label{magnetic}
\end{figure*}

We carried out 2D fully kinetic simulations of collisionless reconnection, with and without a uniform dust background, using the explicit electromagnetic particle-in-cell (PIC) code PSC \citep{Germaschewski2016}.
The simulation domain is periodic in $x$ ($0<x<{{L}_{x}}$), and is bounded by perfectly conducting walls in $z$ at $z=\pm {{L}_{z}}/2$. Here ${{L}_{x}}\times {{L}_{z}}=32{{d}_{i0}}\times 16{{d}_{i0}}$, where ${{d}_{i0}}={{(4\pi {{n}_{i0}}{{e}^{2}}/{{m}_{i}})}^{1/2}}$ is the ion inertial length due to a reference density $n_0$ defined later. Initially, the equilibrium is an antiparallel Harris sheet with the magnetic field
\begin{equation}
    {{B}_{x0}}={{B}_{0}}\tanh (z/\delta ),
\end{equation}
uniform species temperatures $T_\alpha$, and density
\begin{align}
n_{e}= & n_{0}{\rm sech}^{2}\left(z/\delta\right)+n_{b},\\
n_{i}= & n_{0}{\rm sech}^{2}\left(z/\delta\right)+n_{b}+Z_{d}n_{d},
\end{align}
where $n_d$ is the uniform dust background density, $n_0$ is the peak density of the current-carrying portion of the electron and ion species.
The total out-of-plane current density ${{\mathbf{J}}_{0}}=\nabla \times {{\mathbf{B}}_{0}}/{{\mu }_{0}}$ is distributed between the electrons and ions according to ${{J}_{y\alpha }}={{J}_{y}}{{T}_{\alpha }}/(T_e+T_i)$ due to diamagnetic drift. Perturbation is applied to the magnetic field $\delta \mathbf{B}=\hat{\mathbf{z}}\times \nabla \psi $, where
\begin{equation}
    \psi =\delta \psi \cos (2\pi x/{{L}_{x}})\cos (\pi y/{{L}_{y}}).
\end{equation}

To make a reference, we also carried out a run without dust grains, i.e., $n_d=0$. The common parameters of the two runs include $m_i/m_e=100$, $v_{Ae0}/c=0.5$, $n_b/n_0=0.1$, $\delta/d_{i0}=1$, $\delta\psi/(B_0d_{i0})=0.15$, $T_i/T_e=5$. The grid number is $N_x\times N_z=1000\times500$ and the time step size is $\Delta t\omega_{pe0}\approx0.22$. The run with dust also uses $Z_d=1$, $m_d/m_i=16$, $n_d/n_b=5$, $T_d/T_e=1$. Both runs use about $7.5\times10^7$ computational particles per species.

\section{The double Hall pattern}
\label{hall}
The motion of particles in the current sheet can be described by the momentum equations of the three species, respectively
\begin{equation}\label{e}
    \frac{\partial }{\partial t}({{n}_{e}}{{m}_{e}}{{\mathbf{u}}_{e}})+\nabla \cdot ({{n}_{e}}{{m}_{e}}{{\mathbf{u}}_{e}}{{\mathbf{u}}_{e}})=-\nabla \cdot {{\mathbf{P}}_{e}}-e{{n}_{e}}(\mathbf{E}+{{\mathbf{u}}_{e}}\times \mathbf{B})
\end{equation}
\begin{equation}\label{i}
    \frac{\partial }{\partial t}({{n}_{i}}{{m}_{i}}{{\mathbf{u}}_{i}})+\nabla \cdot ({{n}_{i}}{{m}_{i}}{{\mathbf{u}}_{i}}{{\mathbf{u}}_{i}})=-\nabla \cdot {{\mathbf{P}}_{i}}+e{{n}_{i}}(\mathbf{E}+{{\mathbf{u}}_{i}}\times \mathbf{B})
\end{equation}
\begin{equation}\label{d}
    \frac{\partial }{\partial t}({{n}_{d}}{{m}_{d}}{{\mathbf{u}}_{d}})+\nabla \cdot ({{n}_{d}}{{m}_{d}}{{\mathbf{u}}_{d}}{{\mathbf{u}}_{d}})=-\nabla \cdot {{\mathbf{P}}_{d}}-e{{n}_{d}}{{Z}_{d}}(\mathbf{E}+{{\mathbf{u}}_{d}}\times \mathbf{B})
\end{equation}
The spatial characteristic scales thus divide the simulation domain into four regions: the outermost MHD region, where all three species are frozen in the magnetic field lines; the dust diffusion region, where the electrons and ions form charged fluids and are magnetized, with the dust particles demagnetized; the ion diffusion region where only the electron fluid is magnetized and the electron diffusion region where all species are demagnetized. In the electron diffusion region, the motion of the electrons and ions is of high frequency, and the dust species can be seen as an immobile background. From equations (\ref{e}), (\ref{i}), (\ref{d}) combined with the magnetic induction equation, we derived the magnetic field genesis as follows,

\begin{eqnarray}\label{eq:inner}
    &\frac{\partial {{B}_{y}}}{\partial t}=(\mathbf{B}\cdot \nabla )\left( \frac{{{n}_{i}}}{{{n}_{e}}}{{u}_{iy}}-\frac{{{J}_{y}}}{{{n}_{e}}e} \right)-&\\\nonumber&\frac{{{m}_{e}}}{{{n}_{e}}{{e}^{2}}}\nabla \times \left\{ \frac{\partial \mathbf{J}}{\partial t}+\nabla \cdot \left[ (1-\frac{{{n}_{i}}}{{{n}_{e}}}){{n}_{i}}\mathbf{uu}+\frac{{{n}_{i}}}{{{n}_{e}}}(\mathbf{uJ}+\mathbf{Ju})-\frac{\mathbf{JJ}}{{{n}_{e}}e} \right] \right\}\centerdot \hat{\mathbf{y}}
\end{eqnarray}
Here, the Hall term is $-(\mathbf{B}\cdot \nabla )\frac{{{J}_{y}}}{{{n}_{e}}e}$, where $\mathbf{{J}_{y}}=e({{n}_{i}}\mathbf{{u}_{iy}}-{{n}_{e}}\mathbf{{u}_{ey}})$. The term $(\mathbf{B}\cdot \nabla )\frac{n_iu_{iy}}{n_e}$ is due to bulk convection. We denote the term $-\frac{{{m}_{e}}}{{{n}_{e}}{{e}^{2}}}\nabla \times \nabla \cdot \left[ (1-\frac{{{n}_{i}}}{{{n}_{e}}}){{n}_{i}}\mathbf{uu}+\frac{{{n}_{i}}}{{{n}_{e}}}(\mathbf{uJ}+\mathbf{Ju})-\frac{\mathbf{JJ}}{{{n}_{e}}e} \right]\centerdot \hat{\mathbf{y}}$ as the quadratic term.

On the other hand, in the outer region, the motions of the electron fluid and the ion fluid are coupled, serving as a positively charged fluid where the electric charges of the ions are partially screened by the electrons. Effectively, this fluid can be described by
\begin{equation}
    \frac{\partial }{\partial t}({{n}_{+}}{{m}_{+}}{{\mathbf{u}}_{+}})+\nabla \cdot ({{n}_{+}}{{m}_{+}}{{\mathbf{u}}_{+}}{{\mathbf{u}}_{+}})=-\nabla \cdot {{\mathbf{P}}_{+}}+e{{Z}_{+}}{{n}_{+}}(\mathbf{E}+{{\mathbf{u}}_{+}}\times \mathbf{B})
\end{equation}
where ${{n}_{+}}{{m}_{+}}{{\mathbf{u}}_{+}}={{n}_{i}}{{m}_{i}}{{\mathbf{u}}_{i}}+{{n}_{e}}{{m}_{e}}{{\mathbf{u}}_{e}}$, ${{n}_{+}}{{m}_{+}}{{\mathbf{u}}_{+}}{{\mathbf{u}}_{+}}={{n}_{i}}{{m}_{i}}{{\mathbf{u}}_{i}}{{\mathbf{u}}_{i}}+{{n}_{e}}{{m}_{e}}{{\mathbf{u}}_{e}}{{\mathbf{u}}_{e}}$, ${{\mathbf{P}}_{+}}={{\mathbf{P}}_{i}}+{{\mathbf{P}}_{e}}$, $e{{Z}_{+}}{{n}_{+}}(\mathbf{E}+{{\mathbf{u}}_{+}}\times \mathbf{B})=e{{n}_{i}}(\mathbf{E}+{{\mathbf{u}}_{i}}\times \mathbf{B})-e{{n}_{e}}(\mathbf{E}+{{\mathbf{u}}_{e}}\times \mathbf{B})$. In this case, the effective positively charged fluid can be characterized by ${{n}_{+}}\approx {{n}_{i}}$, ${{m}_{+}}\approx {{m}_{i}}$, ${{\mathbf{u}}_{+}}\approx {{\mathbf{u}}_{i}}$, ${{Z}_{+}}=({{n}_{i}}-{{n}_{e}})/{{n}_{i}}$, ${{\mathbf{P}}_{+}}={{\mathbf{P}}_{i}}+{{\mathbf{P}}_{e}}$.

The magnetic field genesis can therefore be derived as
\begin{eqnarray}\label{eq:outer}
   & \frac{\partial {{B}_{y}}}{\partial t}=(\mathbf{B}\cdot \nabla )\left( \frac{{{Z}_{d}}{{n}_{d}}}{{{n}_{+}}}{{u}_{dy}}-\frac{{{J}_{y}}}{{{n}_{+}}e} \right)-&\\\nonumber&
   \frac{{{m}_{+}}}{{{n}_{+}}{{e}^{2}}}\nabla \times \left\{ \frac{\partial \mathbf{J}}{\partial t}+\nabla \cdot \left[ (\frac{{{Z}_{d}}{{n}_{d}}}{{{n}_{+}}}-1){{Z}_{d}}{{n}_{d}}\mathbf{uu}+\frac{{{Z}_{d}}{{n}_{d}}}{{{n}_{+}}}(\mathbf{uJ}+\mathbf{Ju})+\frac{\mathbf{JJ}}{{{n}_{+}}e} \right] \right\}\centerdot \hat{\mathbf{y}}
\end{eqnarray}
the Hall term is $-(\mathbf{B}\cdot \nabla )\frac{{{J}_{y}}}{{{n}_{+}}e}$, where $\mathbf{{J}_{y}}=e({{Z}_{+}}{{n}_{+}}{{\mathbf{u}}_{+y}}-{{Z}_{d}}{{n}_{d}}{{\mathbf{u}}_{dy}})$. We also observe the convective term $(\mathbf{B}\cdot \nabla )\left( \frac{{{Z}_{d}}{{n}_{d}}}{{{n}_{+}}}{{u}_{dy}} \right)$, quadratic term $-\frac{{{m}_{+}}}{{{n}_{+}}{{e}^{2}}}\nabla \times \nabla \cdot \left[ (\frac{{{Z}_{d}}{{n}_{d}}}{{{n}_{+}}}-1){{Z}_{d}}{{n}_{d}}\mathbf{uu}+\frac{{{Z}_{d}}{{n}_{d}}}{{{n}_{+}}}(\mathbf{uJ}+\mathbf{Ju})+\frac{\mathbf{JJ}}{{{n}_{+}}e} \right]\centerdot \hat{\mathbf{y}}$.

\begin{figure}
\centering
	\includegraphics[width=0.99\linewidth]{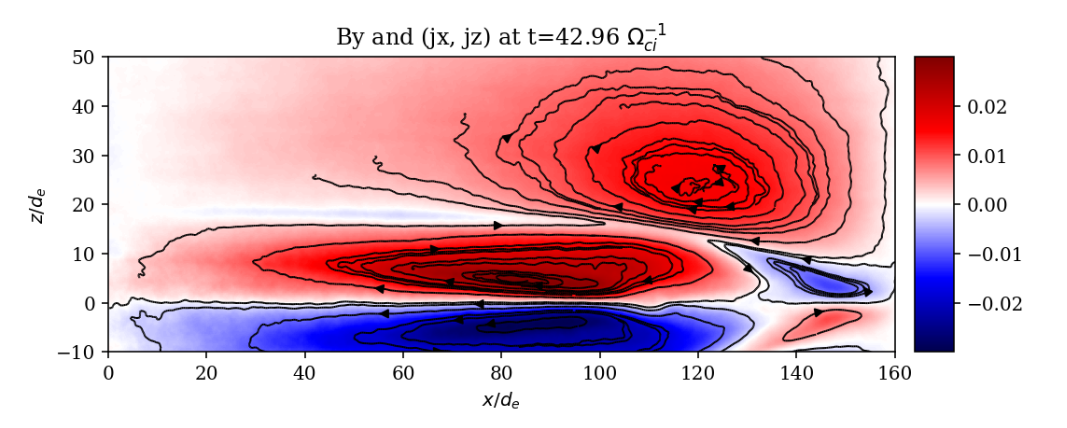}
    \caption{The current closure pattern in collisionless magnetic reconnection with dust grains. We calculated the current density carried by each species by ${{n}_{\alpha }}{{v}_{\alpha }}{{q}_{\alpha }}$, $(\alpha =e,i,d)$ and the total current density $\sum\limits_{\alpha =e,i,d}{{{n}_{\alpha }}{{v}_{\alpha }}{{q}_{\alpha }}}$.}
    \label{closure}
\end{figure}

\begin{figure*}
\centering
\includegraphics[width=1.0\textwidth]{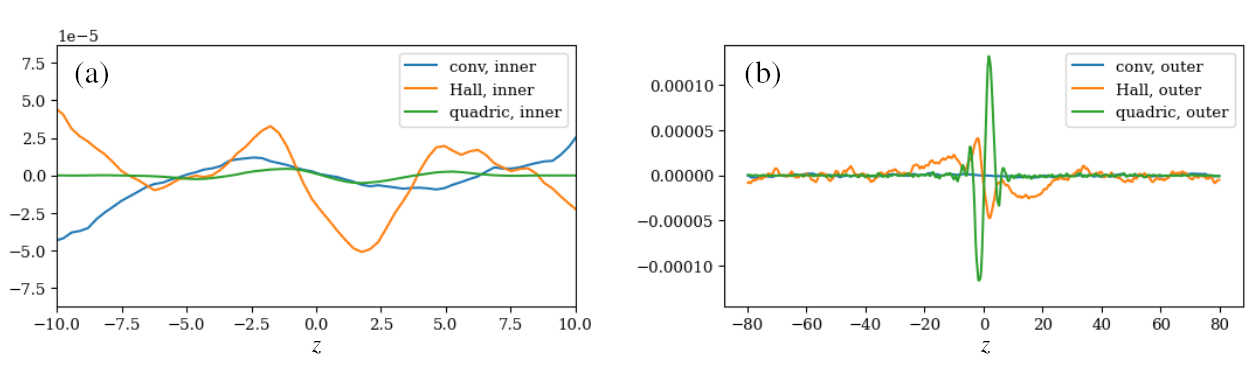}
\caption{The out-of-plane magnetic genesis terms in the inner (left panel) and outer (right panel) regions, where "conv", "Hall", and "quadratic" indicates convection term, Hall term, and quadratic term, respectively, as defined in equations (\ref{eq:inner}) and (\ref{eq:outer}). The data are vertical slices taken at $x=130d_e$ and $t=20/\Omega_{ci}$ from the run with dust grains.}
    \label{terms}
\end{figure*}

\begin{figure*}
\centering
\subfigure[]{\label{5a}\includegraphics[width=0.48\linewidth]{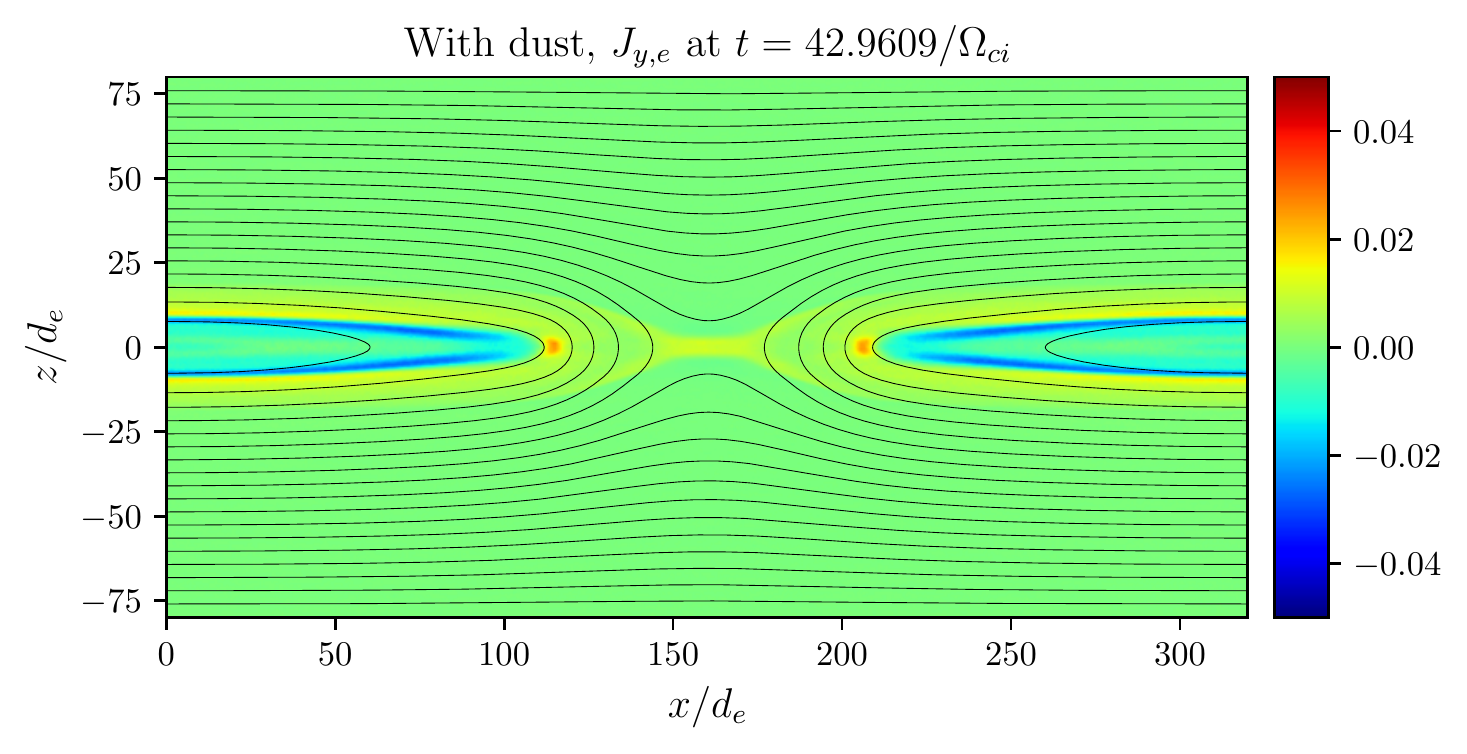}}
\subfigure[]{\label{6a}\includegraphics[width=0.48\linewidth]{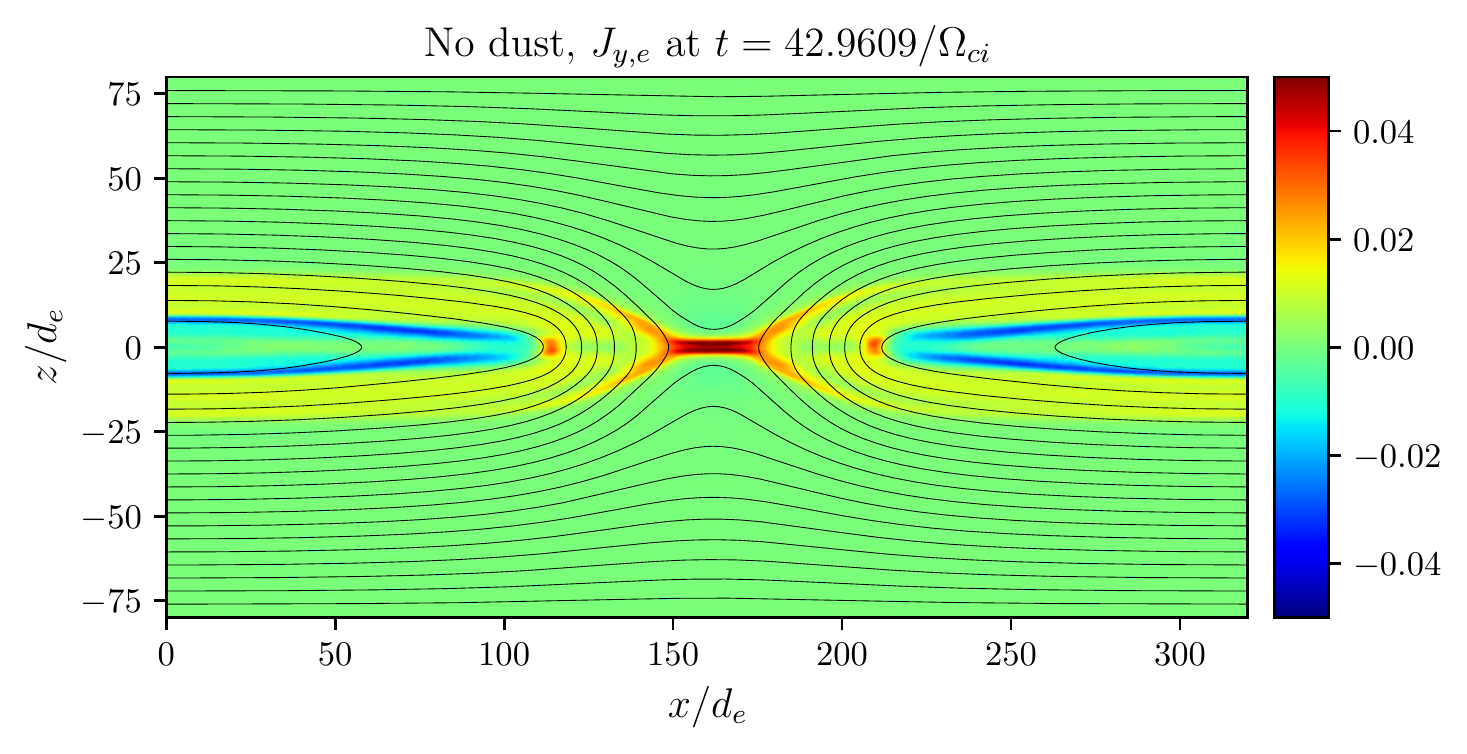}}
\subfigure[]{\label{7a}\includegraphics[width=0.48\linewidth]{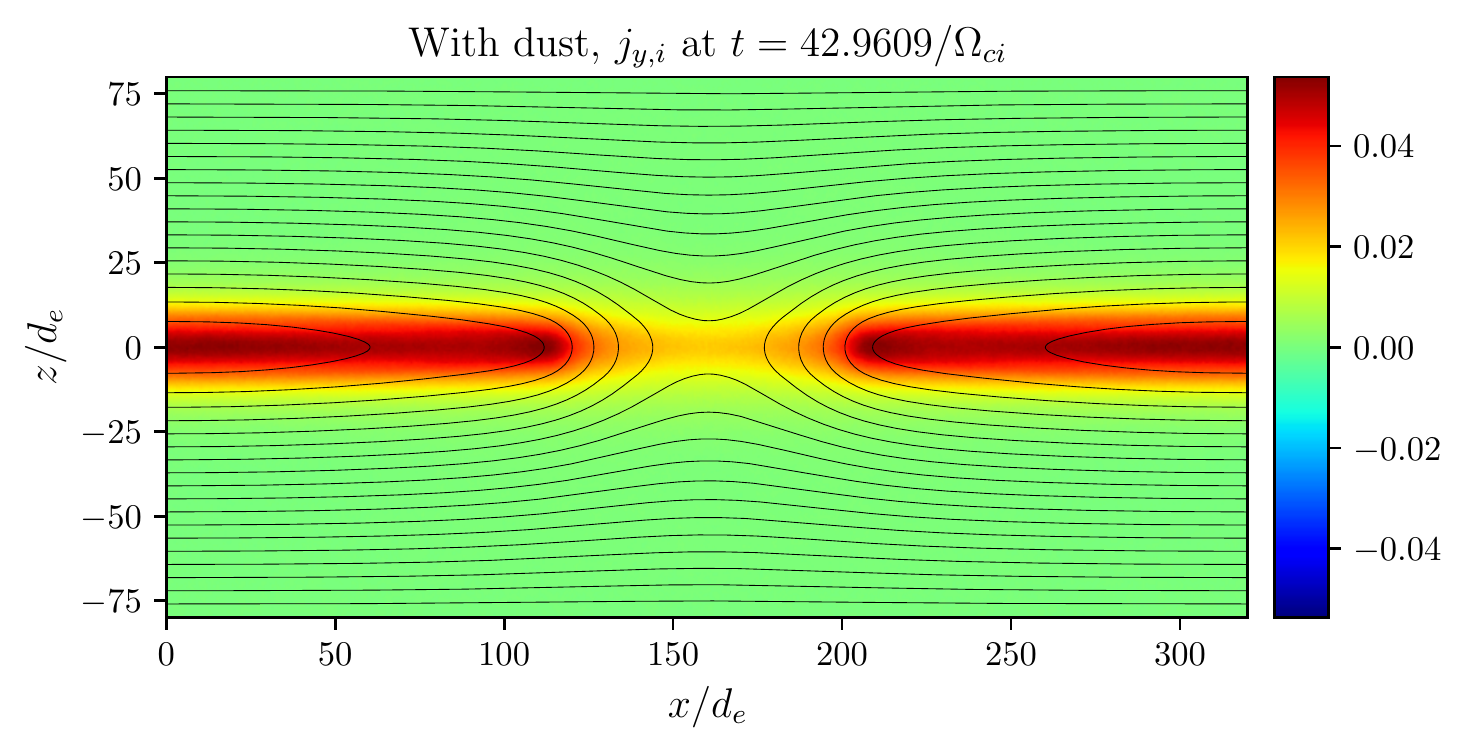}}
\subfigure[]{\label{5b}\includegraphics[width=0.48\linewidth]{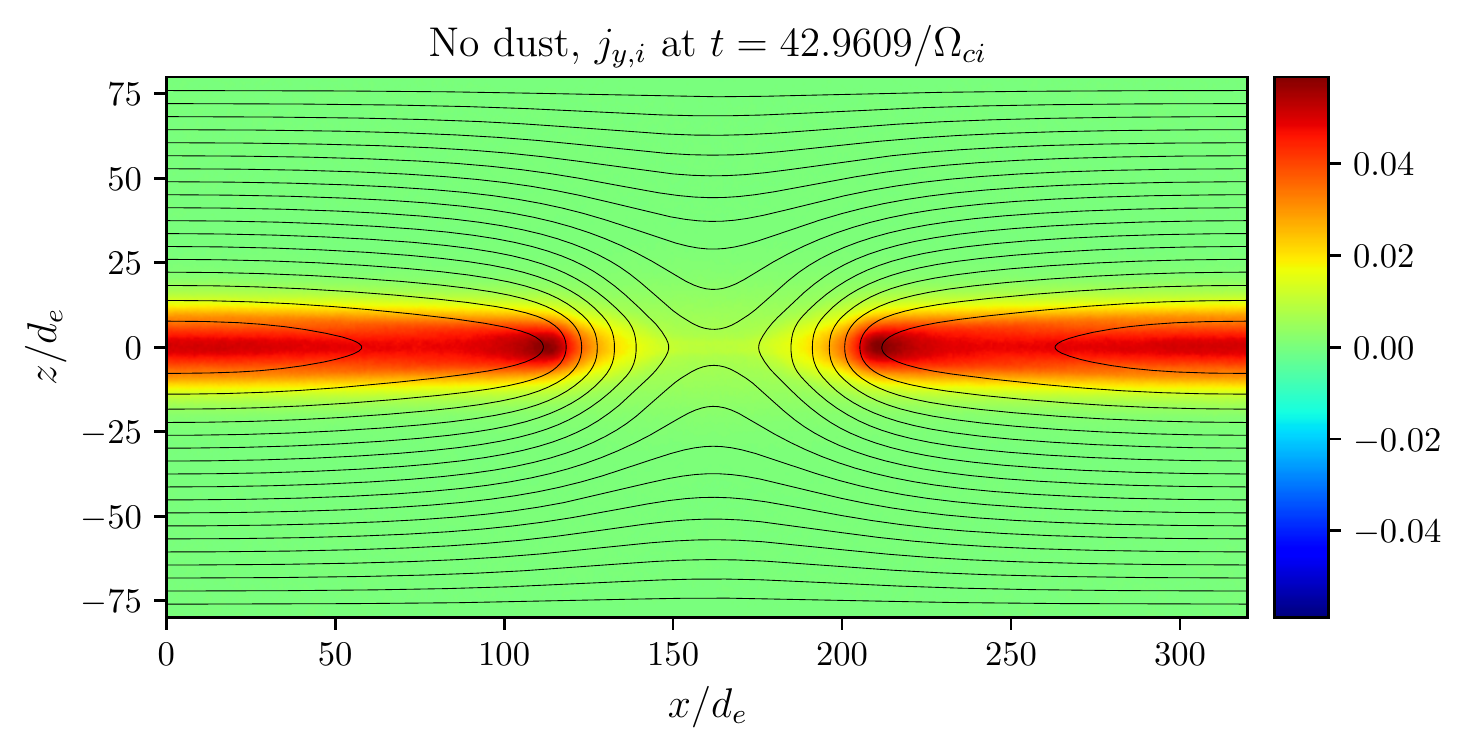}}
\subfigure[]{\label{6b}\includegraphics[width=0.48\linewidth]{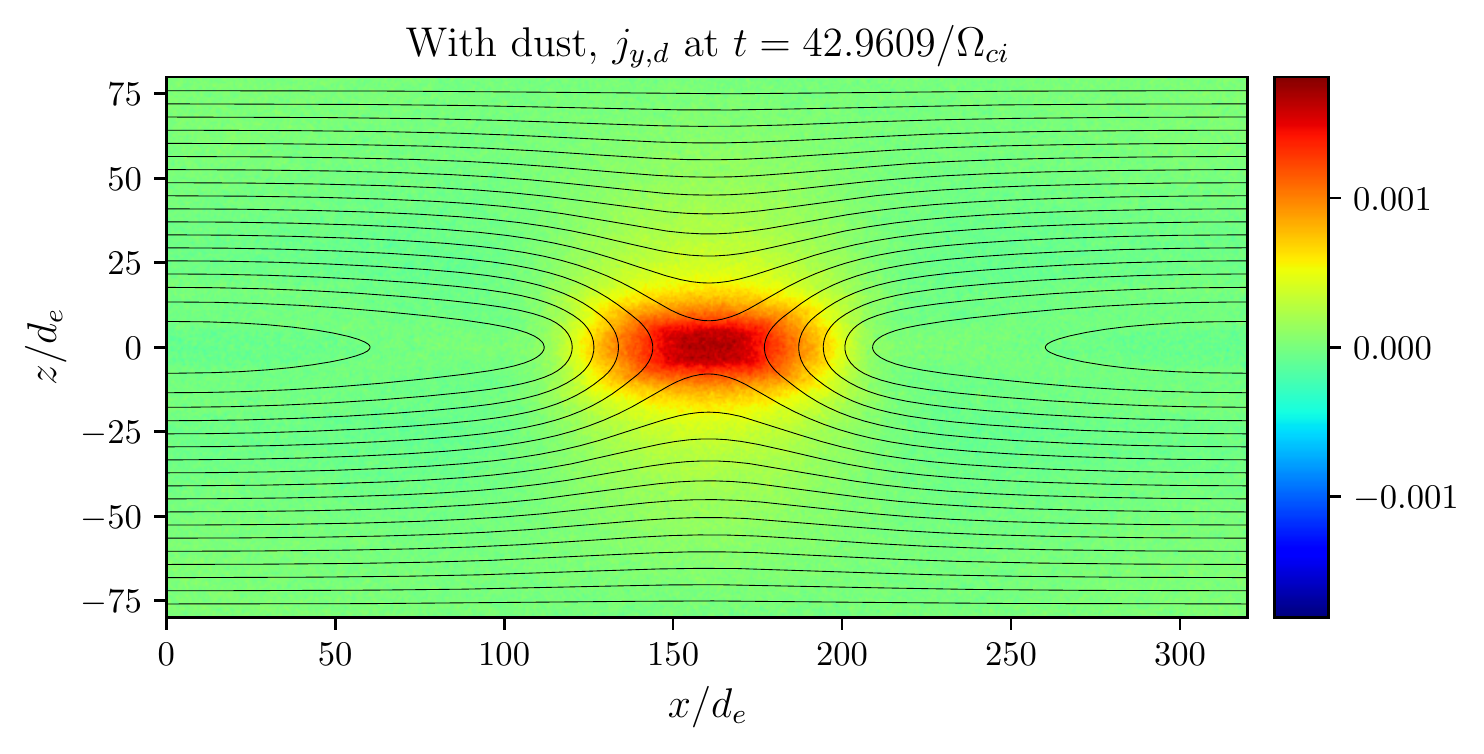}}\hspace{0.015\linewidth}
\subfigure[]{\label{7b}\includegraphics[width=0.4\linewidth]{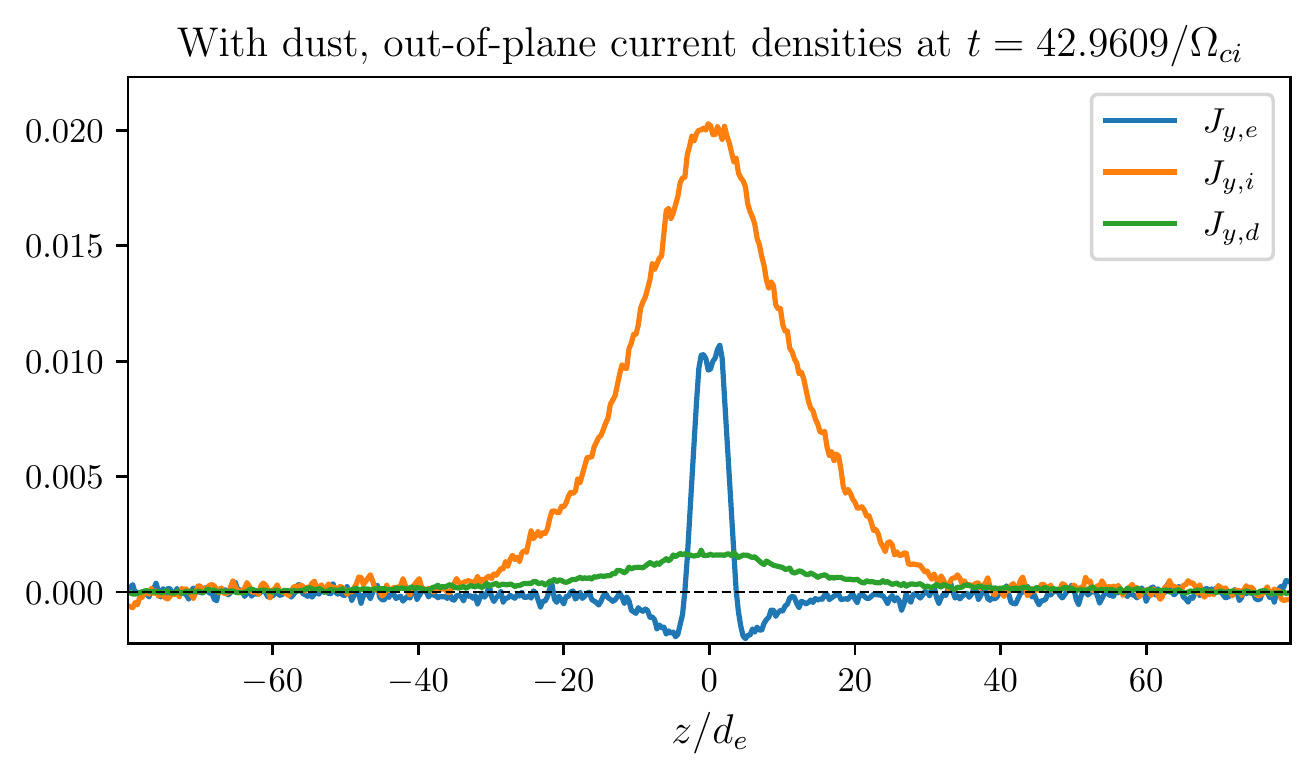}}\hspace{0.065\linewidth}
\caption{Calculated current density of different species from the cases with dust grains and without dust grains.}
\label{current}
\end{figure*}

\begin{figure*}[htbp] 
\centering
\includegraphics[width=7in]{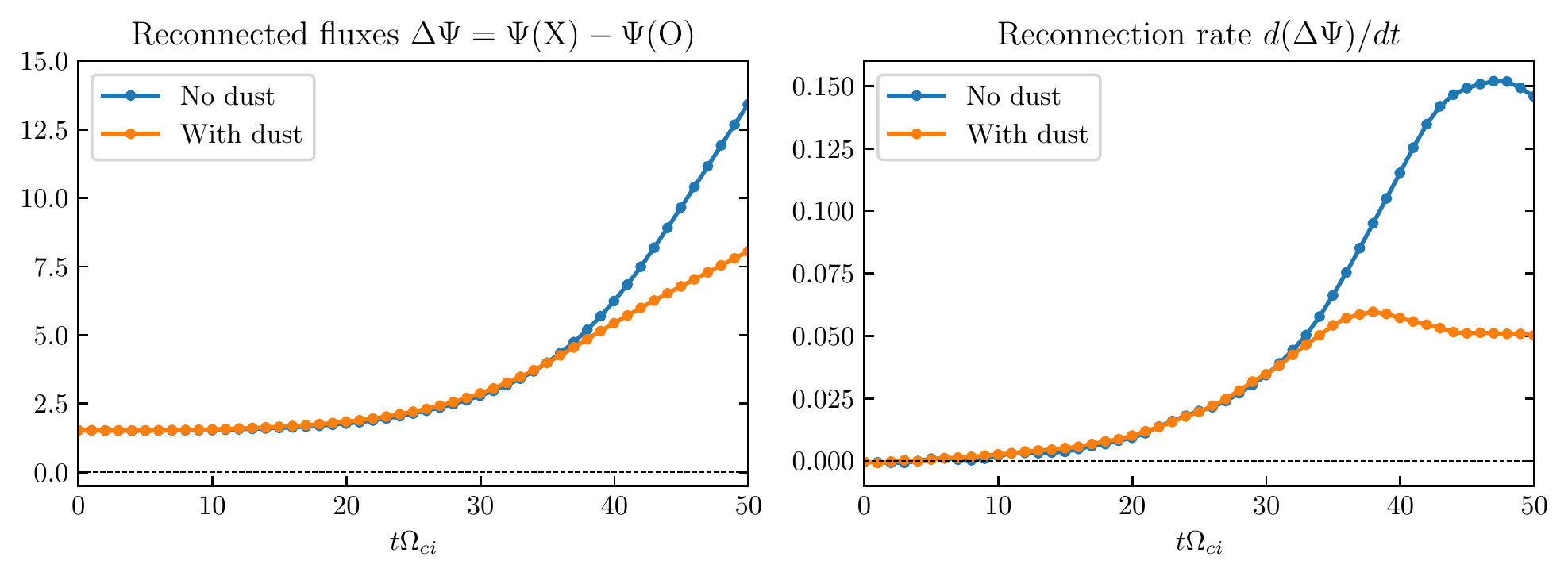} 
\caption{Calculated magnetic reconnection rate in the dusty plasma and the electron-ion plasma.}\label{rate}
\end{figure*}

Due to the mechanism for magnetic field genesis, the most exotic structure emerging in the three-species dusty reconnection is probably the Hall structure. With clear scale separation for ${{d}_{d}}>{{d}_{i}}>{{d}_{e}}$, in-between the electron diffusion region and the ion diffusion region we expect to see a traditional Hall quadruple structure. Going beyond the ion diffusion region while staying inside the dust diffusion region, a reversed Hall structure appears. Fig. \ref{magnetic} shows a time series of the out-of-plane magnetic field ${{B}_{y}}$, where the left panel shows the snapshots at three different time steps for the dusty plasma case and the right panel is the reference run without dust grains. By comparing the ${{B}_{y}}$ patterns between the two columns of Fig. \ref{magnetic}, the case with dust grains clearly presents a double Hall pattern. The outer Hall magnetic fields associated with the negatively charged dust grains exhibit opposite polarity (Fig. \ref{3a}), and their magnitude gradually intensifies with time. Interestingly, the central Hall magnetic field shrinks in Fig. \ref{2b}, accompanied by the thinning of the current sheet and the growing of the outer Hall pattern. The sequence of the emergence of different structures results from the different characteristic time scales associated with the electron, ion, and dust dynamics.

This Hall pattern is formed due to the central current closure pattern in Fig. \ref{closure}, which is the combination of the inner electron-ion current closure and an outer dust current closure. We calculated the current density carried by each species by ${{n}_{\alpha }}{{u}_{\alpha }}{{q}_{\alpha }}$ $(\alpha =e,i,d)$ and the total current $\sum\limits_{\alpha =e,i,d}{{{n}_{\alpha }}{{u}_{\alpha }}{{q}_{\alpha }}}$.

The terms that are responsible for the out-of-plane magnetic field genesis can be seen in Fig. \ref{terms}, where we plotted the different terms in the outer region and the inner region, respectively. From this figure, we could see that Hall terms dominate to form both the inner Hall pattern and the outer Hall pattern.  

Fig. \ref{current} shows the out-of-plane current densities of different species from the two simulations. We plot the current density distribution of the three species at the time when the double Hall magnetic field in the last panel of Fig. \ref{magnetic} is established. The left panel of Fig. \ref{current} shows the current density carried by electrons, ions, and dusty grains, respectively, in the dusty run. For comparison, the top two figures in the right panel depict the electron and ion current density in the dust-free plasma. Clearly, the dust component carries the least electric current density compared with the other two species. The current density carried by electrons, however, is strongly localized around the electron inertial region and manifests a complex structure. We also plot the current decomposition of the three species along $z$ at $x=160 d_e$. Fig. \ref{7b} shows the comparison of current intensity among dust, ion, and electrons. In Fig. \ref{7b}, it is clear that at the moment when the double Hall pattern exists, ion current density is the strongest, electron current sheet manifests a more complex pattern, and dust current density is apparently weaker than the others due to its large mass or, more accurately, its small charge-to-mass ratio.

It can be seen from equation (\ref{eq:inner}) that with fixed ${n}_{i}$, the larger ${n}_{e}$, the stronger the magnetic field genesis. This explains the larger ${B}_{y}$ magnitude in the dust-free run.

\section{Reconnection rate}
\label{Rrate}
Having discussed the exotic double Hall pattern in dusty reconnection, we further look at the reconnection rate in dusty plasmas. The reconnection rate is a defining characteristic of the process, depicting the rate at which the magnetic topology changes and the energy conversion occurs. The presence of dust grains inevitably influences the reconnection rate. The reconnected flux and reconnection rate are presented in Fig. \ref{rate} for comparison, where the reconnection rate is normalized over the ion Alfv\'{e}n time ${{v}_{Ai}}$. Fig. \ref{rate} demonstrates clearly that the presence of dust grains severely slows down the reconnection process. It is conceivable that in a plasma that is undergoing fast reconnection, if dust grains get accumulated simultaneously, fast reconnection could be suspended, and the procedure becomes more complex.

The importance of dust species in shaping the relevant processes can be estimated by the characteristic spatial and temporal scales. The larger the difference between the dust timescale and the ion timescale, the slower that dust comes into play. In the limit that the dust is too heavy that it serves as immobile background, it will have little influence in the relevant processes.

\begin{figure*}
\centering
\subfigure[]{\label{fig:motion-with-dust}\includegraphics[width=0.48\textwidth]{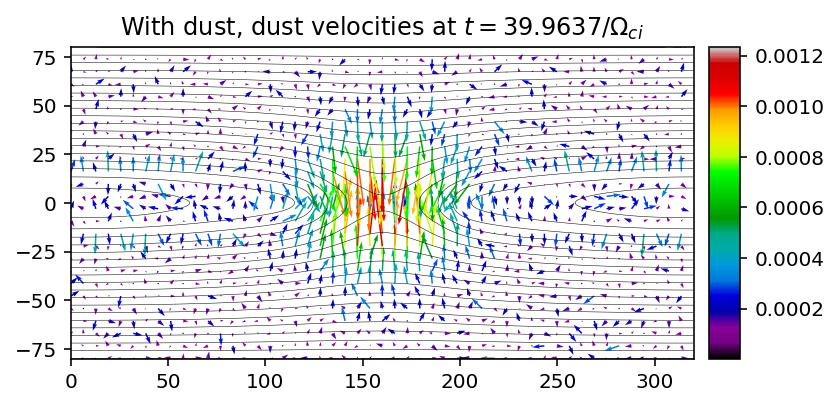}}
\subfigure[]{\label{fig:motion-without-dust}\includegraphics[width=0.48\textwidth]{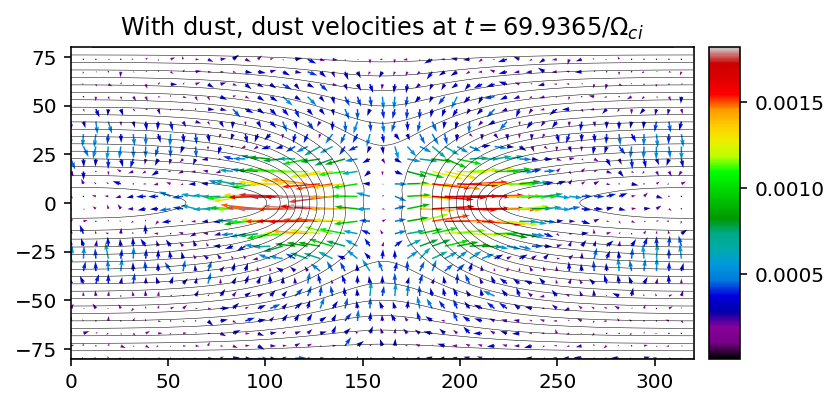}}
\caption{Motion of dust grains at two different time steps. The arrow directions represent the local velocity direction. The arrow lengths and colors are scaled by the local dust grain speed. The underlying thin gray lines are magnetic field lines.}
\label{dustm}
\end{figure*}

We also looked into the motion of the dust during the reconnection process. In Fig. \ref{dustm}, the dust grains gradually concentrate to the center of the current sheet. As the dust motion becomes pronounced, the dusty plasma reconnection rate starts to deviate from the dust-free reconnection rate (see Fig. \ref{rate}). The concentration of dust grains in the center may also serve as a potential mechanism for dusts to grow in size during planetary formation \citep{liu2020}, where the electromagnetic field may help the dust grains to accumulate further. This is a highly nonlinear process and thus will not be explored here. As a result of reconnection, the dust grains are finally convected out of the current sheet by the outflow.

\section{Conclusion}
\label{sum}
In conclusion, we reported the discovery of a novel double Hall pattern associated with collisionless reconnection in dusty plasmas. This pattern is composed of a traditional Hall magnetic field and a reversed Hall magnetic field, characterized by the inertial scale of electron, ion, and dust grain, respectively. This is unique for dusty plasmas, due to alternating charge signs and clear scale separation where all three species participate in the dynamics. We also presented the current density distribution in dusty plasmas, where the current density carried by dust grains, though weaker than the electron and ion current density, plays a role in shaping relevant structures. In particular, the reconnection rate of a dusty plasma is slower than that of an electron-ion plasma with other identical parameters, indicating that dust formation may slow down the reconnection process. The scenario discussed in this study may be realized and provide reference to the dusty plasma experiments like Laboratory for Plasma Nanosynthesis (LPN) and MDPX \citep{Merlino1998}.

In the future, we may further explore the scaling laws of dust reconnection with respect to different mass ratios ${{m}_{d}}/{{m}_{i}}$, different concentrations of dust ${{n}_{d}}$ and different charged state ${{Z}_{d}}$, as well as investigate the different regimes in dusty plasma reconnection. We could also study the coupling and interactions between different scales. The double Hall effect on dusty plasma turbulent energy spectra may also lead to exotic phenomena. These open questions remain to be explored.

\section*{Acknowledgements}

C.D. was supported by the U.S. Department of Energy under contract number DE-AC02-09CH11466.

\section*{Data Availability}

The PSC simulation code is available at \url{https://github.com/psc-code/psc}. The data that support the findings of this study are available on request from the corresponding author.



\bibliographystyle{mnras}
\providecommand{\noopsort}[1]{}\providecommand{\singleletter}[1]{#1}%

\label{lastpage}
\end{document}